\documentclass[fleqn,usenatbib]{mnras}

\usepackage{newtxtext,newtxmath}

\usepackage[T1]{fontenc}

\DeclareRobustCommand{\VAN}[3]{#2}
\let\VANthebibliography\thebibliography
\def\thebibliography{\DeclareRobustCommand{\VAN}[3]{##3}\VANthebibliography}

\usepackage{graphicx}	
\usepackage{amsmath}	

\title{Star Formation in the NE-Circinus Molecular Cloud Complex}

\author[D. Rose]{D. Rose\thanks{E-mail: danrose74@gmail.com}
\\
Independent Researcher\\
}

\date{Accepted 2026 June 23. Received 2026 June 23; in original form 2026 May 4}

\pubyear{\the\year{}}

\begin{document}
\label{firstpage}
\pagerange{\pageref{firstpage}--\pageref{lastpage}}
\maketitle


\begin{abstract}
This paper presents the first dedicated characterisation of the NE-Circinus 
Complex (TGU~H1984, DCld~320.7$-$03.6), a previously unstudied star-forming dark cloud complex serendipitously identified in archival \textit{Herschel} SPIRE observations targeting the foreground Bok globules BHR~99 and BHR~100. 
Using \textit{Herschel} SPIRE photometry, Planck Galactic Cold Clumps data, 
near-infrared 2MASS colour excess mapping, AllWISE photometry, Gaia~DR3 
photometry, and a three-dimensional dust extinction map, the complex and its 
embedded YSO population are characterised for the first time. Two independent 
photometric methods place the complex at $750 \pm 50$~pc. Three morphologically 
distinct components are identified: a dense main cloud body, a diffuse eastern 
component, and a northern filamentary extension. The main cloud body has a 
SPIRE-derived dust temperature of $14.5 \pm 0.5$~K. Near-infrared $H-K$ colour 
excess mapping yields a core gas mass of ${\sim}277~\mathrm{M}_\odot$ and a 
total gas mass of ${\sim}439~\mathrm{M}_\odot$, consistent with the Planck PGCC 
column density. An AllWISE YSO census identifies 
39 candidates across all three components, confirming active star formation 
throughout the complex. Multi-epoch NEOWISE-R photometry reveals three 
significantly variable sources, including one aperiodic dipper --- a previously 
uncatalogued YSO exhibiting dimming consistent with inner-disc occultation. Two 
previously uncatalogued compact Bok globules are identified on the western edge 
of the northern extension, both detected in SPIRE continuum emission. That this 
complex went uncharacterised despite \textit{Herschel} data being publicly 
available since 2013 underscores the scientific value of archival examination 
and the incomplete state of southern sky molecular cloud inventories.
\end{abstract}




\section{Introduction}

The southern sky hosts numerous poorly studied dark cloud complexes that represent 
important but undercharacterised sites of low-mass star formation. Many of these 
structures were catalogued in the \citet{dobashi2005} Throughfall Galaxy Universe 
(TGU) extinction atlas and in the \citet{Lynds1962} catalogue of dark nebulae and the 
DC catalogue of southern dark clouds \citep{Hartley1986}, yet the 
majority have received no dedicated multiwavelength follow-up beyond their initial 
photographic identification. Heterogeneous survey coverage across the southern sky 
means that while some clouds have been characterised at far-infrared, millimetre, 
and radio wavelengths, others remain known only from their optical extinction 
signatures. Understanding the physical properties and embedded stellar populations 
of such clouds is essential for building a complete census of star formation in the 
solar neighbourhood.

The cloud complex studied in this paper --- catalogued as TGU~H1984 \citep{dobashi2005} and 
DCld~320.7$-$03.6 \citep{Hartley1986}, hereafter the NE-Circinus Complex --- lies at Galactic 
coordinates approximately $l = 320.7\degr$, $b = -3.4\degr$ to $-3.8\degr$, at 
an estimated distance of $750 \pm 50$~pc (Section~\ref{sec:distance}). The complex 
was serendipitously captured in \textit{Herschel} Space Observatory SPIRE 
observations originally targeting the foreground Bok globules BHR~99 and BHR~100 
\citep{Bourke1995}, which are physically unassociated foreground objects at 
${\sim}350$~pc \citep{Bourke1995}. Despite its angular extent of approximately 
$20 \times 15$~arcmin in the SPIRE 250~\micron\ band, and the detection of multiple 
cold clumps by the Planck Galactic Cold Clumps (PGCC) catalogue 
\citep{PlanckCollaboration2016}, the NE-Circinus Complex has received no dedicated 
study in the published literature.

The most recent comprehensive study of the CMC region, \citet{Kerr2025}, 
characterised the stellar populations of the Circinus Complex using \textit{Gaia} 
astrometry and identified multiple subgroups associated with the CMC gas clouds. The 
\textit{Herschel} observations used here were obtained under programme OT2\_tbourke\_3 
(PI: T.~Bourke). \citet{Sadavoy2018} used the same SPIRE data in a multi-cloud dust 
emission study but produced only temperature and optical depth maps for the foreground 
globules; no source-specific science results for the NE-Circinus Complex were reported. 
The complex has therefore gone entirely uncharacterised in far-infrared emission beyond 
its appearance in survey catalogues.

The NE-Circinus Complex lies approximately $2.5\degr$ north-east of Cir-E and $3.8\degr$ north-east of Cir-W in projection. Although spatially and morphologically distinct from those well-studied clouds, the complex may represent a physically connected but spatially removed 
component of the broader Circinus molecular cloud region. Faint diffuse extinction is visible in DSS2 Red imaging of the intervening region, suggesting a possible continuity of dust structure between the NE-Circinus Complex and the main Circinus clouds.

In this paper the first dedicated characterisation of the NE-Circinus 
Complex is presented. The following datasets were used: \textit{Herschel} SPIRE photometry of the main cloud body, Planck PGCC catalogued properties for all three components, AllWISE photometry for YSO classification using the \citet{Koenig2014} two-colour scheme, and \textit{Gaia}~DR3 
stellar photometry combined with the \citet{Guo2021} three-dimensional dust extinction 
map for distance determination.

The paper is structured as follows. Section~\ref{sec:observations} describes the 
observational data and archival resources employed. Section~\ref{sec:distance} 
presents our distance determination. Section~\ref{sec:results} presents our results. 
Section~\ref{sec:discussion} discusses the physical properties of the complex and its 
likely connection to the CMC. Section~\ref{sec:conclusions} summarises our conclusions.

\section{OBSERVATIONS}
\label{sec:observations}

\subsection{Herschel SPIRE}
\label{sec:spire}

Far-infrared observations of the NE-Circinus Complex field were obtained with the 
\textit{Herschel} Space Observatory \citep{Pilbratt2010} as part of Open Time 
programme OT2\_tbourke\_3 (PI: T.~Bourke, Lonely Cores: Star Formation in 
Isolation). The programme targeted the foreground Bok globules BHR~99 and BHR~100 
\citep{Bourke1995}. Two SPIRE \citep{Griffin2010} large-scan observations were 
obtained on March 1, 2013: observation 1342265318, centred on BHR~100, and 
observation 1342265319, centred on BHR~99. Each observation was executed in 
LargeScan mode with a duration of 480~s, producing simultaneous maps in the PSW 
(250~\micron), PMW (350~\micron), and PLW (500~\micron) bands. The two observations 
were combined into a mosaic using the standard \textit{Herschel} Interactive 
Processing Environment (HIPE) pipeline, processed to Level~2.5, covering both 
foreground globules and an extended region to their east. It was during processing 
of this mosaic that the main body of the NE-Circinus Complex was identified as a 
bright, spatially extended far-infrared source; its subsequent investigation forms 
the basis of the present paper.

The SPIRE mosaic covers the main cloud body essentially in full. The eastern 
component lies entirely outside the SPIRE footprint, and the northern filamentary 
extension has only its southernmost portion within the covered area. Beam FWHMs 
are 18.1, 24.9, and 36.6~arcsec at 250, 350, and 500~\micron\ respectively, with 
pixel scales of 6, 10, and 14~arcsec~pixel$^{-1}$. Point-source-calibrated (\textsc{psrc}) maps for compact source photometry and the extended-source-calibrated (\textsc{extd}) maps for diffuse cloud emission analysis were used.

\subsection{AllWISE}
\label{sec:allwise}

Mid-infrared photometry was drawn from the AllWISE Source Catalogue 
\citep{Wright2010, Cutri2013}, queried via the VizieR TAP service. Separate cone 
searches were performed centred on each of the three cloud components: the main 
cloud body (radius 15~arcmin), the eastern component (radius 15~arcmin), and the 
northern filamentary extension (radius 20~arcmin), with BHR~99 and BHR~100 excluded 
by applying a positional exclusion radius of $0.05\degr$ around each globule.

YSO candidates were identified and classified using the two-colour scheme of 
\citet{Koenig2014}, applied in \textsc{topcat} to each cone search catalogue. The 
scheme uses the $W1-W2$ versus $W2-W3$ colour--colour diagram to separate Class~I, 
Class~I/flat-spectrum, and Class~II sources, with explicit exclusion of likely 
contaminants including AGB stars, Classical Be stars, and star-forming galaxies.

The $W3$ band (12~\micron) is sensitive to diffuse polycyclic aromatic hydrocarbon 
(PAH) emission, which is particularly apparent in the AllWISE images of the eastern 
component; YSO counts for this component should therefore be treated with slight 
caution.

\subsection{NEOWISE-R Multi-Epoch Photometry}
\label{sec:neowise}

Multi-epoch $W1$ (3.4~\micron) and $W2$ (4.6~\micron) photometry was retrieved for 
the YSO candidates identified in Section~\ref{sec:allwise} from the NEOWISE-R 
Single Exposure (L1b) Source Table \citep{Mainzer2014} via the IRSA TAP service, 
using a 3~arcsec positional match radius. Raw single-exposure data were binned into 
10-day windows and quality-filtered (\texttt{qi\_fact} $> 0$, \texttt{saa\_sep} $> 
5$) to produce cleaned per-epoch mean magnitudes.

All candidates were cross-matched against SIMBAD to identify known contaminants. 
Two sources satisfying the \citet{Koenig2014} Class~II colour cuts were identified 
as Mira variables: IRAS~15259$-$6059 and OGLE~GD-LPV-8662. Both sources were removed from the sample.

\subsection{Planck PGCC}
\label{sec:pgcc}

Cold clump properties for regions outside the SPIRE footprint were taken from the 
Planck Galactic Cold Clumps catalogue \citep[PGCC;][]{PlanckCollaboration2016}, 
queried via the VizieR TAP service using the table \texttt{J/A+A/594/A28/pgcc}. 
Three PGCC sources fall within the extent of the cloud complex 
(Table~\ref{tab:pgcc}). PGCC source detection used the 857~GHz Planck map as the 
detection image, with physical characterisation --- including dust temperature and 
column density --- derived at lower frequencies where the beam FWHM is approximately 
4.5~arcmin. The resulting temperatures and column densities therefore represent 
beam-averaged quantities over each clump's angular extent, a point of interpretive 
importance when comparing with the higher-resolution SPIRE results discussed in 
Section~\ref{sec:discussion_mass}.

\begin{table}
\centering
  \caption{Planck Galactic Cold Clumps \citep{PlanckCollaboration2016} detected
    within the extent of the NE-Circinus Complex, queried from the VizieR
    catalogue \texttt{J/A+A/594/A28/pgcc}. Dust temperatures and column densities
    are beam-averaged quantities over the Planck beam ($\sim$4.5~arcmin FWHM).
    No reliable temperature fit is available for G320.96$-$3.81.
    Clump gas masses are computed at the adopted distance of 750~pc.}
  \label{tab:pgcc}
  \begin{tabular}{lccc}
    \hline
   Name & $T_\mathrm{dust}$ & $N(\mathrm{H_2})$ & Component \\
         & (K) & (cm$^{-2}$) & \\
    \hline
    G320.67$-$3.67 & $13.7 \pm 0.9$ & $2.36 \times 10^{21}$ & Main body \\
    G320.96$-$3.81 & --- & $6.4 \times 10^{20}$ & Eastern \\
    G320.81$-$3.23 & $15.9 \pm 3.1$ & $3.80 \times 10^{20}$ & Northern \\
    \hline
  \end{tabular}
\end{table}

\subsection{Distance Determination}
\label{sec:distmethod}

\subsubsection{Gaia DR3}
\label{sec:gaia}

Stellar photometry for distance determination was drawn from the \textit{Gaia} Data 
Release~3 catalogue \citep{GaiaCollaboration2023} via cone searches performed toward 
the main cloud body, the eastern component, and the northern filamentary extension, 
as well as a control field at $l = 320.8\degr$, $b = -6.35\degr$. From the retrieved 
photometry, $B_{\rm P}-R_{\rm P}$ colour was plotted against geometric distance for 
all stars in each field, producing colour-distance diagrams in which a sharp reddening 
onset marks the near face of an intervening dust layer.

\subsubsection{Guo et al.\ (2021) 3D Dust Map}
\label{sec:dustmap}

Three-dimensional dust extinction profiles were extracted along sightlines toward each 
cloud component and the control field at $l = 320.8\degr$, $b = -6.35\degr$, using 
the southern sky extinction map of \citet{Guo2021}, queried at the HEALPix pixel 
positions corresponding to each sightline.

\subsection{Near-Infrared Extinction Mapping}
\label{sec:nirext}
 
Near-infrared $H-K$ colour excess mapping was used to characterise the
line-of-sight extinction toward all three components of the
NE-Circinus Complex.
Near-infrared photometry was drawn from the Two Micron All Sky Survey
\citep[2MASS;][]{Skrutskie2006} Point Source Catalogue, queried via the
VizieR TAP service (table \texttt{II/246/out}).
Cone searches were performed centred on each component: the main cloud
body (RA\,$231.767\degr$, Dec\,$-61.025\degr$) at radii of $0.1\degr$
(6~arcmin) and $0.167\degr$ (10~arcmin); the eastern component
(RA\,$232.8\degr$, Dec\,$-61.5\degr$) at $0.1\degr$; and the northern
filamentary extension (RA\,$231.35\degr$, Dec\,$-60.78\degr$) at
$0.333\degr$ (20~arcmin, encompassing the full projected length of the
extension).
A control field centred at $l = 320.8\degr$, $b = -6.35\degr$
(RA\,$231.97\degr$, Dec\,$-63.58\degr$) was queried at $0.25\degr$
radius to establish the intrinsic stellar $H-K$ colour baseline.
Only sources with photometric quality flag \texttt{AAA} in all three
bands were retained.
 
To isolate stars physically background to the cloud, each catalogue was
cross-matched against \textit{Gaia}~DR3 \citep{GaiaCollaboration2023}
by sky position within 1~arcsec.
Geometric distances were estimated as $d = 1000/\varpi$~pc, where
$\varpi$ is the Gaia DR3 parallax in milliarcseconds.
Sources with $d > 750$~pc and parallax signal-to-noise
$\varpi/\sigma_\varpi > 5$ were retained as background stars.
Sources deviating by more than $3\sigma$ from the sample mean $H-K$
were excluded as likely spurious detections.
The resulting background star counts are summarised in
Table~\ref{tab:extinction}.
 
The mean $H-K$ colour of the background-filtered control field,
$\langle H-K \rangle_0 = 0.116 \pm 0.085$~mag (2676~stars), was
adopted as the intrinsic stellar colour baseline.
The colour excess $E(H-K) = \langle H-K \rangle_\mathrm{cloud} -
\langle H-K \rangle_0$ was converted to visual extinction using
$A_V = E(H-K)/0.063$ \citep{Cardelli1989}, and subsequently to
molecular hydrogen column density using
$N(\mathrm{H_2}) = 9.4 \times 10^{20}\,A_V$~cm$^{-2}$
\citep{Bohlin1978}.
 
Gas masses for the main cloud body were computed as:
\begin{equation}
  M_\mathrm{gas} = N(\mathrm{H_2})\, \mu\, m_\mathrm{H}\, \pi r^2
  \label{eq:mass}
\end{equation}
where $\mu = 2.8$ is the mean molecular weight per hydrogen molecule,
$m_\mathrm{H}$ is the hydrogen atom mass, and $r$ is the physical
aperture radius at 750~pc.
Two apertures were used for the main cloud body: a 6~arcmin aperture
(physical radius 1.3~pc) targeting the dense core encompassing the
Planck PGCC clump G320.67$-$3.67, and a 10~arcmin aperture
(physical radius 2.2~pc) approximately encompassing the full extent of
detectable cold dust emission in the SPIRE 250~$\mu$m map.
Although the cloud extends to approximately 15~arcmin in optical
extinction, the mean $H-K$ excess becomes increasingly diluted by
low-column peripheral material at larger radii, rendering mass estimates
beyond $\sim$10~arcmin unreliable.
 
Mass estimates were not derived for the eastern component or the
northern filamentary extension.
For the eastern component, the extinction signal is weak
($E(H-K) = 0.031$~mag, $A_V = 0.49$~mag), consistent with the diffuse,
low-column nature of the structure, and the large fractional uncertainty
makes any mass estimate unreliable.
For the northern filamentary extension, the elongated filamentary
geometry is poorly matched by a circular aperture, and the mean
extinction over the 20~arcmin search area is heavily diluted by the
large fraction of low-extinction sky enclosed; a reliable total mass
cannot be derived without resolved extinction mapping along the
filament.
The $H-K$ excess and column density for both components are reported as
characterisation measurements only (Table~\ref{tab:extinction}).

\section{DISTANCE}
\label{sec:distance}

\begin{figure*}
    \centering
    \includegraphics[width=\textwidth]{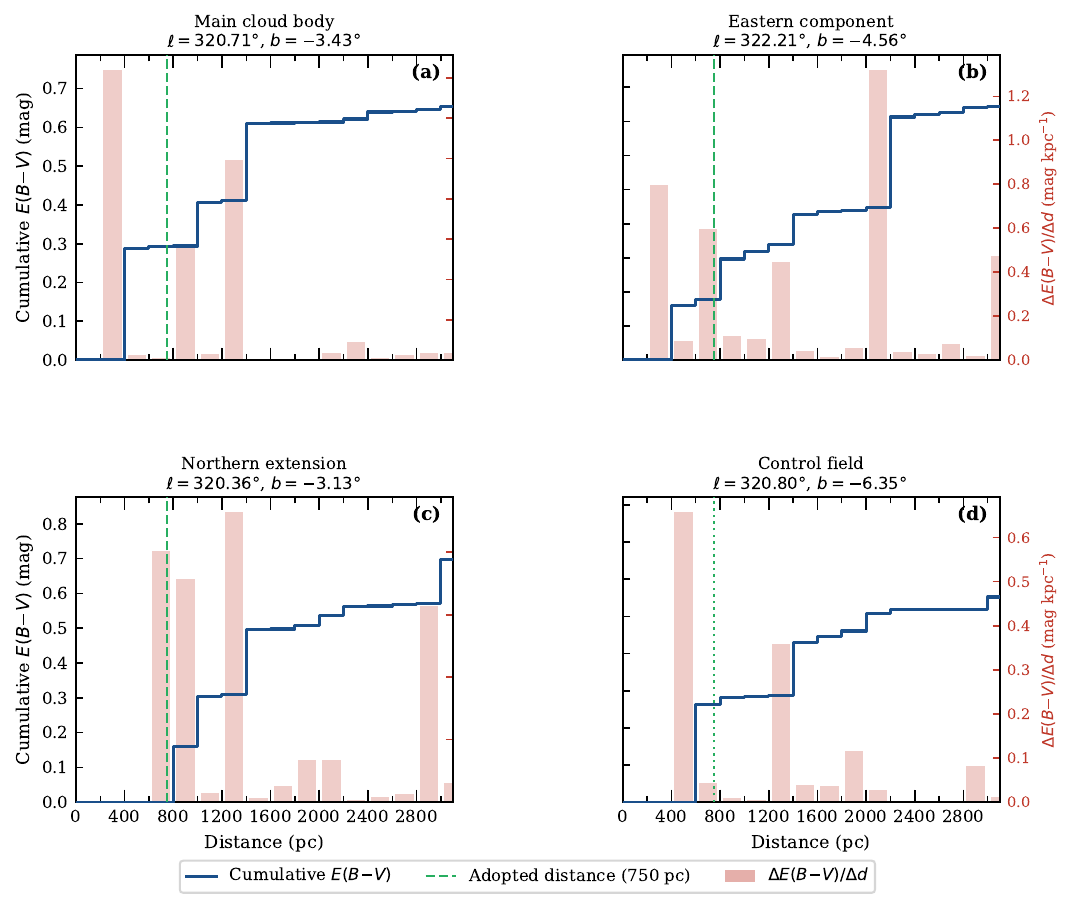}
    \caption{Guo et al. (2021) three-dimensional dust extinction 
    profiles toward the three components of the NE-Circinus Complex 
    and a control field at $\ell = 320.8\degr$, $b = -6.35\degr$. 
    The line plot shows the cumulative $E(B-V)$ as a function of 
    distance; column bars show the extinction gradient 
    $\Delta E(B-V)/\Delta d$ in each 200\,pc bin, centred on the 
    interval to which they correspond. The dashed green line marks 
    the adopted distance of 750\,pc (dotted in the control field 
    panel for reference). A foreground Galactic dust layer at 
    $\sim$300\,pc produces a pronounced extinction step in the main 
    cloud body, eastern component, and control field sightlines; 
    this component is unrelated to the NE-Circinus Complex. A 
    second step at $\sim$750\,pc is present in all three cloud 
    sightlines but absent in the control field, confirming the 
    physical association of the dust with the complex at the 
    adopted distance. The northern extension sightline shows no 
    foreground step at $\sim$300\,pc; the reason for this 
    difference is unclear.}
    \label{fig:guo_dust}
\end{figure*}

Three-dimensional dust extinction profiles were extracted along 
sightlines toward each cloud component and a control field at 
$\ell = 320.8\degr$, $b = -6.35\degr$, using the southern sky 
extinction map of \citet{Guo2021}. For each sightline the map 
provides cumulative $E(B-V)$ in 200\,pc bins out to 6\,kpc; 
the extinction gradient $\Delta E(B-V)/\Delta d$ per bin 
highlights where extinction is being deposited most rapidly 
along the line of sight. Both quantities are shown in 
Figure~\ref{fig:guo_dust}.

A pronounced extinction step, visible as both a jump in cumulative 
$E(B-V)$ and elevated gradient bars, is present in the range 
600--800\,pc in all three cloud sightlines but absent in the 
control field at $b = -6.35\degr$, confirming that the dust is 
physically associated with the NE-Circinus Complex at the adopted 
distance. A foreground Galactic dust layer at $\sim$300\,pc is 
visible in the main cloud body, eastern component, and control 
field sightlines; it is unrelated to the complex. Secondary 
features at $\sim$1200\,pc in the main cloud body and eastern 
component sightlines likely trace the outer envelope of the 
Circinus Molecular Cloud.

The northern extension sightline shows no foreground dust layer 
at $\sim$300\,pc, in contrast to the main cloud body, eastern 
component, and control field sightlines. This is consistent with 
the more complex foreground dust structure at lower Galactic 
latitudes ($b > -3\degr$) where the foreground sheet becomes 
patchy or falls below the map detection threshold. Despite the 
absence of a foreground signal, the northern extension shows a 
clear extinction step at $\sim$750\,pc --- visible as both 
elevated gradient bars and a jump in cumulative $E(B-V)$ --- 
consistent with the adopted distance and confirming physical 
association with the NE-Circinus Complex.

The adopted distance of $750 \pm 50$\,pc is further corroborated 
by two coincident stellar populations identified by 
\citet{Kerr2025}: the Theia~861 co-moving group 
\citep{Kounkel2019} at 716\,pc projected against the eastern 
component, and the OC-0625 stellar subgroup at 821\,pc coincident 
with the main cloud body. These bracket our adopted distance from 
below and above respectively. In the absence of molecular line 
observations providing a systemic radial velocity for the gas, 
neither population can be kinematically confirmed as physically 
associated with the cloud, but their distances are mutually 
consistent with $750 \pm 50$\,pc. At this distance, 
1\,arcmin corresponds to 0.22\,pc.

\section{RESULTS}
\label{sec:results}

\subsection{Morphology}
\label{sec:morphology}

\subsubsection{Main Cloud Body}
\label{sec:maincloud}

Optical morphology was assessed using DSS2 Red imaging retrieved from the ESO 
Online Digitized Sky Survey\footnote{\url{http://archive.eso.org/dss/dss}}; the 
three components of the NE-Circinus Complex are shown in 
Figure~\ref{fig:overview}. In DSS2 Red imaging, the main cloud body presents as 
an amorphous dark cloud with well-defined western and southern boundaries, beyond 
which the stellar background reasserts sharply. Toward the east the cloud thins 
progressively, with discrete regions of enhanced extinction gradually dissolving 
into the surrounding stellar field rather than terminating at a defined edge. 
Several discrete regions of enhanced extinction are visible within the cloud body, 
particularly concentrated along the southern edge and western boundary, connected 
by more diffuse obscuring material.

The \textit{Herschel} SPIRE 250~\micron\ emission traces a similar complex, 
filamentary structure with multiple local maxima, confirming the presence of cold 
dust throughout the cloud body (Figure~\ref{fig:spire}). The brightest 
250~\micron\ emission peaks are spatially coincident with the regions of strongest 
optical extinction in the DSS2 imaging. The SPIRE mosaic covers the main cloud 
body essentially in full, with the 250~\micron\ map spanning approximately $20 
\times 15$~arcmin at this component's extent, corresponding to a physical size of 
approximately $4.4 \times 3.3$~pc at 750~pc.

\subsubsection{Eastern Component}
\label{sec:eastern}

The eastern component is considerably more diffuse than the main cloud body, 
presenting only subtly in DSS2 Red imaging as a region of marginally reduced 
stellar density. It never clearly emerges as a distinct identifiable structure in 
optical imaging, appearing instead as a slight suppression of the background 
stellar field with a centrally located region of marginally enhanced obscuration. 
Its connection to the main cloud body to its west is suggested by the optical 
imaging but cannot be confirmed from DSS2 data alone.

The eastern component lies entirely outside the SPIRE footprint and therefore has 
no \textit{Herschel} far-infrared coverage. Its reality as a distinct cloud 
structure is confirmed by the detection of one PGCC cold clump (G320.96$-$3.81) 
and by the presence of an embedded YSO population identified from AllWISE 
photometry (Section~\ref{sec:yso}). Diffuse PAH emission is apparent in AllWISE 
$W3$ images of this component, consistent with the presence of embedded or nearby 
young stars heating surrounding material.

\subsubsection{Northern Filamentary Extension}
\label{sec:northern}

\begin{figure*}
    \centering
    \includegraphics[width=0.78\textwidth]{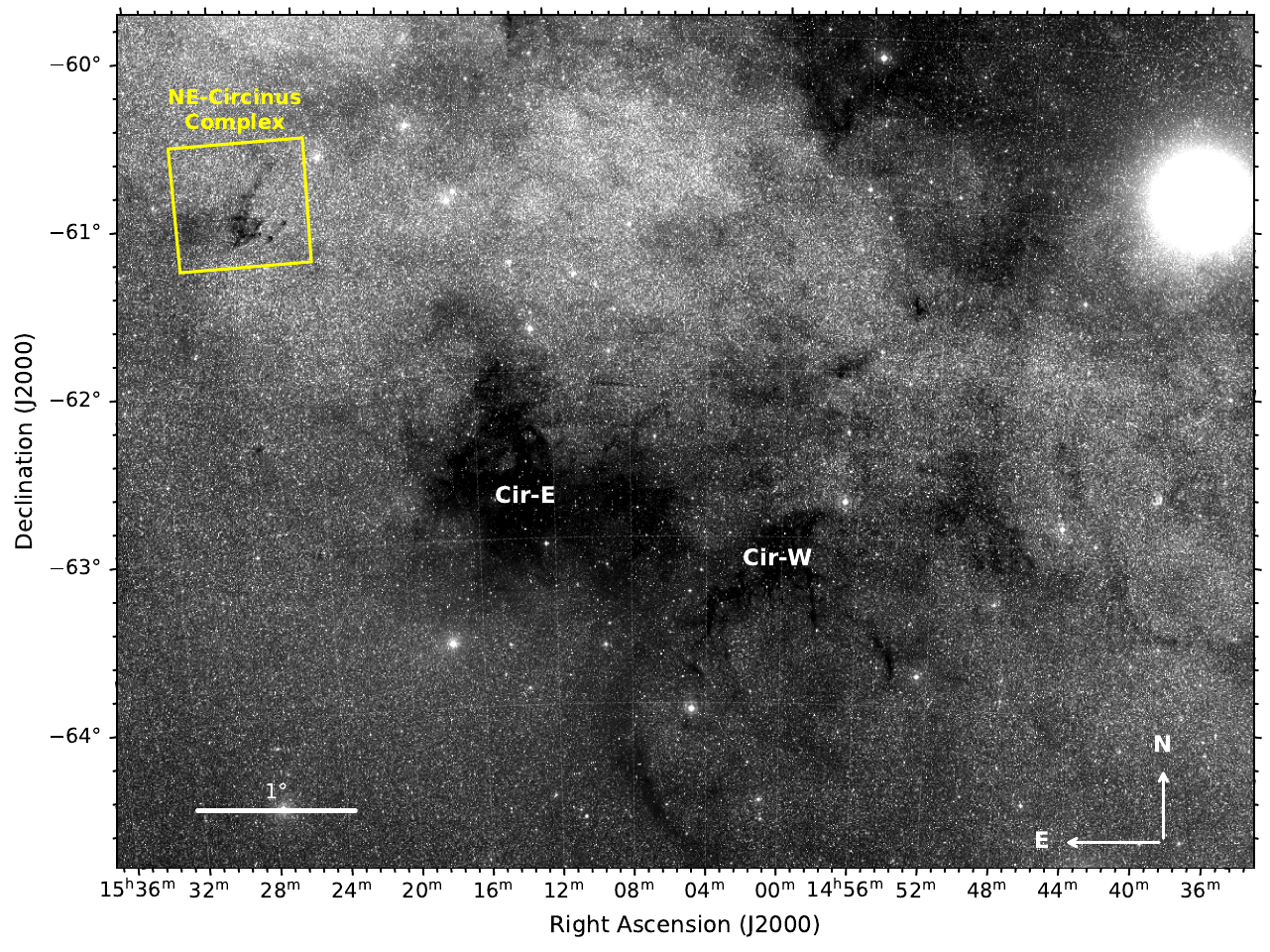}
    \vspace{2pt}
    \includegraphics[width=0.8\textwidth]{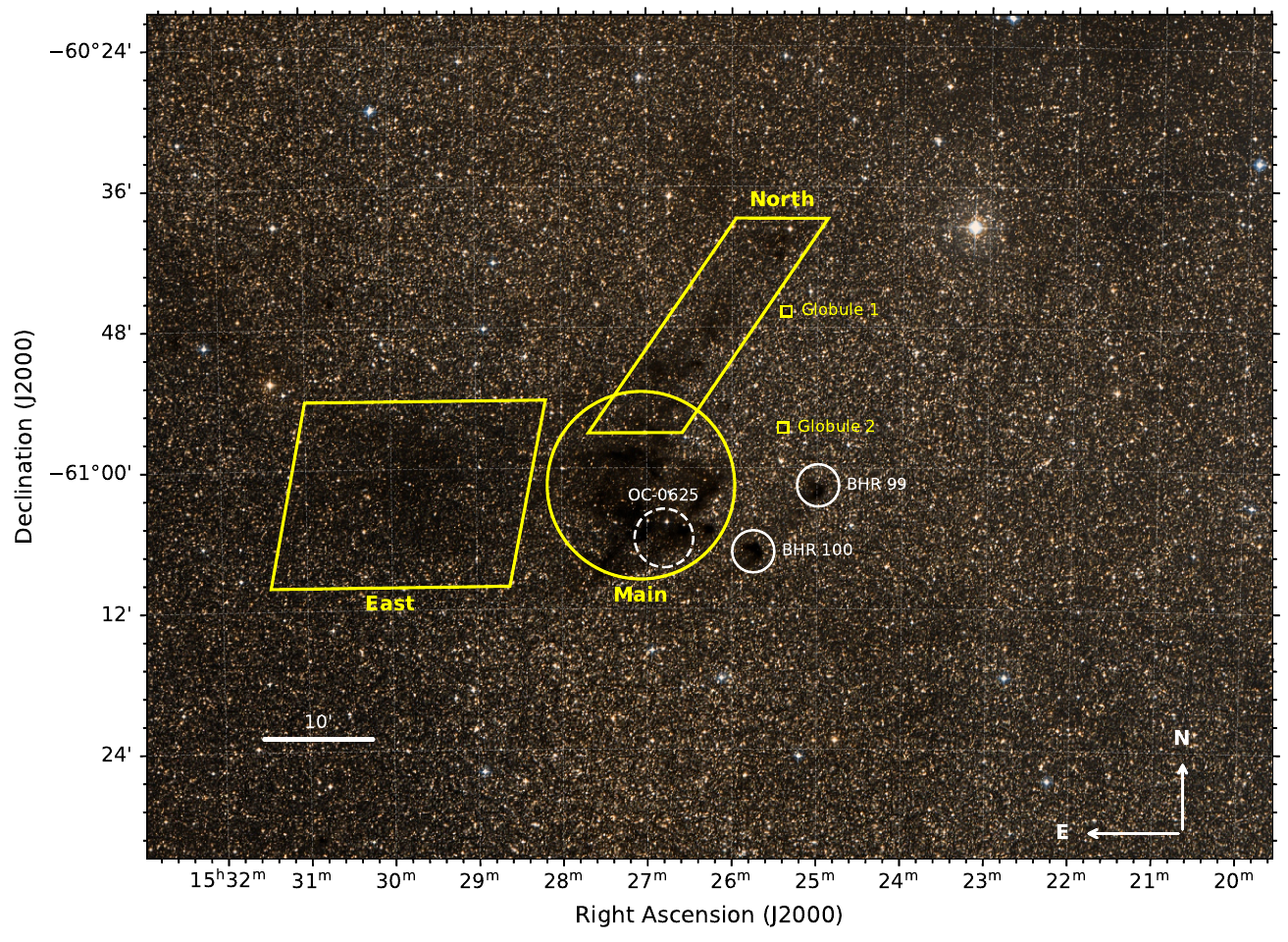}
    \caption{Multi-wavelength imaging of the NE-Circinus Complex and its 
    relation to the broader Circinus star-forming region. (a) Wide-field 
    DSS2 greyscale image (6.8$^\circ \times 5.1^\circ$) showing the 
    NE-Circinus Complex (yellow box) in relation to the Circinus East 
    (Cir-E) and Circinus West (Cir-W) molecular clouds. (b) Close-up DSS2 
    colour image showing the three morphological components of the 
    NE-Circinus Complex: the main cloud body (circle), northern filamentary 
    extension (North), and eastern component (East). The dashed circle marks 
    the mean sky position of the OC-0625 stellar subgroup \citep{Kerr2025}. 
    The Bok globules BHR~99 and BHR~100 are identified to the west of the 
    main cloud body. Newly identified Bok globules NE-Cir Globule~1 and 
    Globule~2 are marked with squares. In both panels, north is up and east 
    is left.}
    \label{fig:overview}
\end{figure*}

\begin{figure*}
    \centering
    \includegraphics[width=\textwidth]{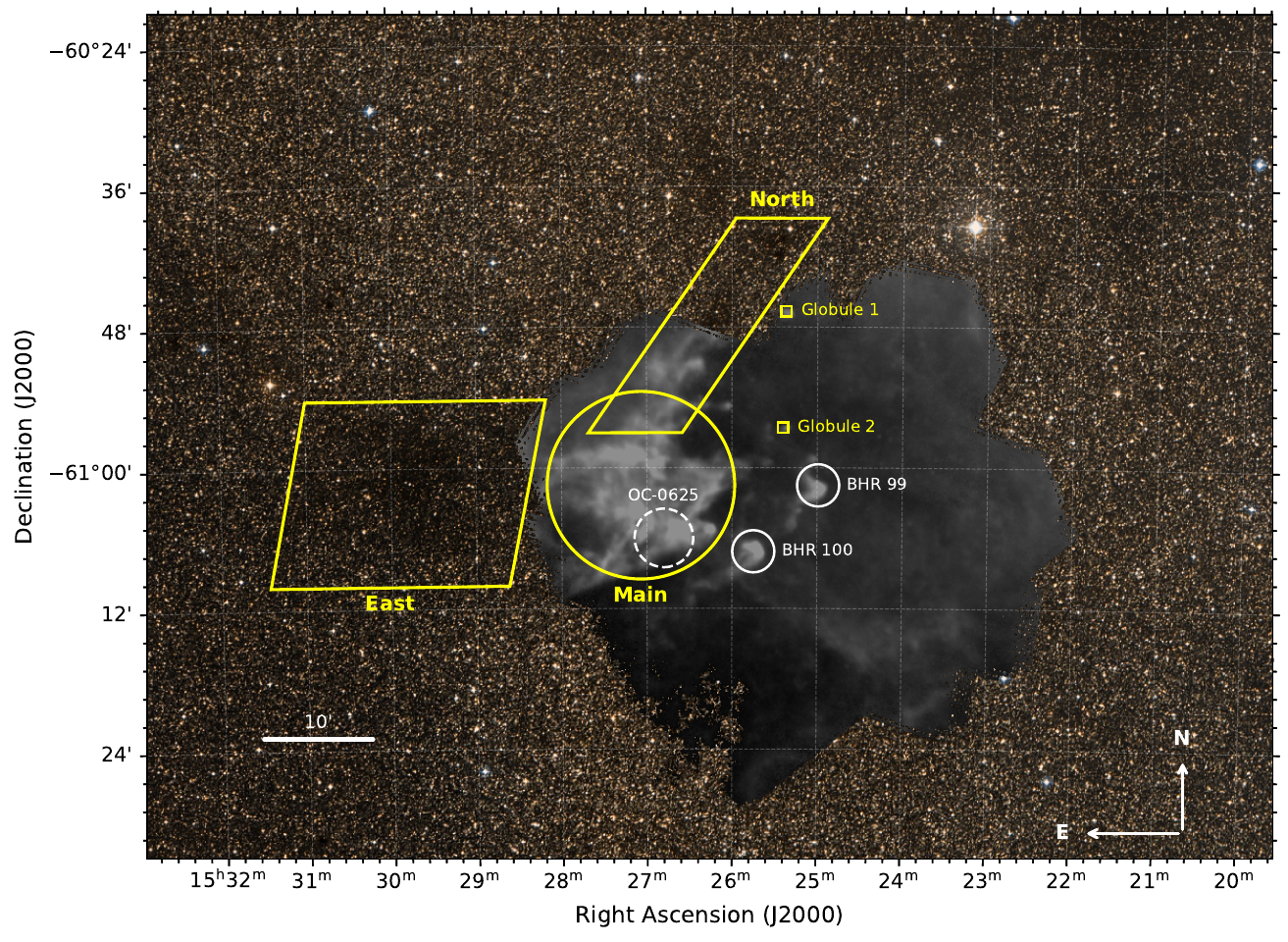}
    \caption{The same field as Figure~\ref{fig:overview}b showing a 
    composite of the DSS2 optical image and the Herschel SPIRE 250~$\mu$m 
    emission (greyscale), revealing the far-infrared dust structure of the 
    complex. The SPIRE mosaic boundary is visible as the sharp edge in the 
    far-infrared emission; the eastern component lies entirely outside the 
    SPIRE footprint. Component boundaries and source labels are as in 
    Figure~\ref{fig:overview}b. North is up and east is left.}
    \label{fig:spire}
\end{figure*}

The northern filamentary extension projects northward from the main cloud body as 
an easily identifiable structure in DSS2 Red imaging 
(Figure~\ref{fig:overview}b). In its southern portion it presents as a dense, 
clumpy chain of discrete, well-defined extinction features. Moving northward, the 
structure gradually thins and becomes more diffuse, meandering north-westward 
across the sky before dissolving progressively into the background stellar field. 
The extension has a projected length of approximately 16~arcmin, corresponding to 
approximately 3.5~pc at 750~pc.

Two compact, apparently uncatalogued Bok globules are visible off the western edge 
of the extension (Table~\ref{tab:globules}). Neither appears in SIMBAD or major 
southern dark cloud catalogues and both warrant future dedicated study.

Only the southernmost portion of the northern extension falls within the SPIRE 
footprint, with the majority of the structure lying outside the \textit{Herschel} 
coverage. As with the eastern component, its physical reality is confirmed by PGCC 
cold clump detections and an embedded YSO population 
(Section~\ref{sec:yso}).

\subsection{Dust Properties and Gas Mass}
\label{sec:dust}
 
\subsubsection{Main Cloud Body}
\label{sec:dust_main}
 
\begin{figure}
    \includegraphics[width=\columnwidth]{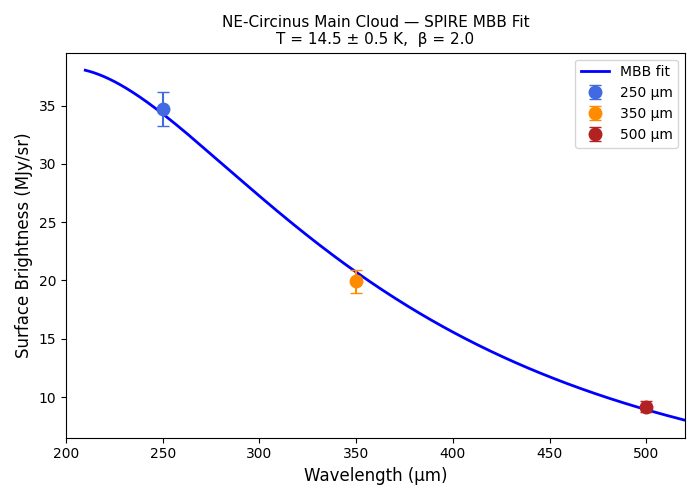}
    \caption{Modified blackbody fit to the \textit{Herschel} SPIRE 250, 350,
    and 500~$\mu$m surface brightnesses of the NE-Circinus main cloud body,
    measured within an 8~arcmin radius aperture centred at
    RA\,$=231.767\degr$, Dec\,$=-61.025\degr$.
    The fit uses $\beta = 2.0$ fixed and yields
    $T_\mathrm{dust} = 14.5 \pm 0.5$~K.
    Error bars include the 4~per~cent SPIRE absolute flux calibration
    uncertainty \citep{Griffin2010} added in quadrature with the
    background scatter.}
    \label{fig:sed}
\end{figure}

The dust temperature of the main cloud body was determined by fitting a
single-temperature modified blackbody of the form
$S_\nu \propto \nu^\beta B_\nu(T)$ to the SPIRE 250, 350, and 500~$\mu$m
flux densities, with the emissivity index fixed at $\beta = 2$ and dust
opacity $\kappa_{300} = 0.08$~m$^2$~kg$^{-1}$
\citep{OssenkopfHenning1994}.
Aperture photometry was performed on the extended-source-calibrated
SPIRE maps using a circular aperture of 8~arcmin radius centred on the
main cloud body at RA\,$231.767\degr$, Dec\,$-61.025\degr$, with
background subtracted using the median of three off-cloud positions. The resulting SED and best-fit modified blackbody are shown in Figure~\ref{fig:sed}.
This yields a dust temperature of $T_\mathrm{dust} = 14.5 \pm 0.5$~K.
 
Although a pixel-by-pixel temperature and column density map is preferable, it could not be robustly produced from the SPIRE data in this field. The zero-point correction method 
of \citet{Sadavoy2018} was investigated but found to be 
inapplicable at this Galactic latitude ($b \approx -3.4\degr$): 
the NE-Circinus Complex contributes less than 1\,MJy\,sr$^{-1}$ 
of far-infrared emission, while the underlying Galactic 
background at Planck resolution exceeds $10^4$\,MJy\,sr$^{-1}$, 
making the cloud signal unresolvable from the Galactic background 
at Planck beam scales. Alternative approaches designed for 
large-scale Galactic plane surveys require continuous multi-degree 
coverage unavailable for this small targeted mosaic. The 
integrated modified blackbody fit presented here uses local 
background subtraction from three off-cloud reference positions, 
which mitigates the zero-point uncertainty for the temperature 
determination; however, absolute column density maps from SPIRE 
remain unreliable without a reliable zero-point correction, and 
mass characterisation therefore relies on the independent 
near-infrared extinction method \citep{Cardelli1989, Bohlin1978}.

The gas mass of the main cloud body was determined from near-infrared
$H-K$ colour excess mapping of background stars (Section~\ref{sec:nirext}).
Results are summarised in Table~\ref{tab:extinction}.
For the 6~arcmin aperture centred on the dense core,
$E(H-K) = 0.154 \pm 0.007$~mag corresponds to $A_V = 2.44$~mag and
$N(\mathrm{H_2}) = 2.3 \times 10^{21}$~cm$^{-2}$, giving a core gas
mass of $277^{+38}_{-36}$~M$_\odot$ at 750~pc.
For the 10~arcmin aperture encompassing the full cloud extent,
$E(H-K) = 0.088 \pm 0.004$~mag gives a total gas mass of
$439^{+60}_{-57}$~M$_\odot$ at 750~pc.
The quoted uncertainties reflect the distance uncertainty of $\pm$50\,pc 
(a fractional uncertainty of $\sim$13 per cent on mass, since $M \propto 
d^2$). Systematic uncertainties in the assumed extinction law 
\citep{Cardelli1989} and the $N({\rm H_2})/A_V$ conversion 
\citep{Bohlin1978} contribute an additional $\sim$20--30 per cent in 
quadrature, reflecting the known variation in $R_V$ across diffuse and 
dense cloud environments. The total uncertainty on the derived masses 
is therefore of order 30--40 per cent, and the quoted values should be 
interpreted accordingly.
 
The column density derived from the 6~arcmin aperture,
$N(\mathrm{H_2}) = 2.3 \times 10^{21}$~cm$^{-2}$, is in close
agreement with the Planck PGCC value for G320.67$-$3.67 of
$2.36 \times 10^{21}$~cm$^{-2}$ (Section~\ref{sec:pgcc}), providing
independent confirmation of the extinction measurement.
The PGCC clump mass of $\sim$105~M$_\odot$ is smaller than both
extinction-derived masses, as expected: the Planck beam at
$\sim$4.5~arcmin FWHM samples only the dense clump identified by
Planck source extraction, whereas the extinction apertures integrate
over the full column including diffuse envelope material.

\begin{table*}
  \caption{Near-infrared $H-K$ colour excess measurements and derived
    properties for all three components of the NE-Circinus Complex,
    from background-filtered 2MASS photometry
    (Section~\ref{sec:nirext}).
    Masses are quoted at the adopted distance of 750~pc; the range in
    parentheses reflects the $\pm50$~pc distance uncertainty.
    No reliable mass is derived for the eastern component or northern
    extension (see text).}
  \label{tab:extinction}
  \begin{tabular}{llccccc}
    \hline
    Component & Aperture & Stars & $E(H-K)$ & $A_V$ & $N(\mathrm{H_2})$ & $M_\mathrm{gas}$ \\
              &          &       & (mag)    & (mag) & (cm$^{-2}$)       & (M$_\odot$) \\
    \hline
    Main body  & 6~arcmin  & 287 & $0.154\pm0.007$ & 2.44 & $2.3\times10^{21}$ & 277 (241--315) \\
    Main body  & 10~arcmin & 1503 & $0.088\pm0.004$ & 1.39 & $1.3\times10^{21}$ & 439 (382--499) \\
    Eastern    & 6~arcmin  & 447 & $0.031\pm0.004$ & 0.49 & $4.6\times10^{20}$ & — \\
    Northern   & 20~arcmin & 537 & $0.087\pm0.004$ & 1.38 & $1.3\times10^{21}$ & — \\
    \hline
  \end{tabular}
\end{table*}
 
\subsubsection{Eastern Component}
\label{sec:dust_eastern}
 
The eastern component lies entirely outside the SPIRE footprint and has
no \textit{Herschel} far-infrared coverage.
Near-infrared extinction mapping yields a weak colour excess of
$E(H-K) = 0.031 \pm 0.004$~mag ($A_V = 0.49$~mag,
$N(\mathrm{H_2}) = 4.6 \times 10^{20}$~cm$^{-2}$), consistent with
the diffuse, low-column nature of this component.
One Planck PGCC cold clump is detected in this component: G320.96$-$3.81,
for which no reliable temperature or column density fit is available
in the catalogue.
 
\subsubsection{Northern Filamentary Extension}
\label{sec:dust_northern}
 
The northern filamentary extension has only its southernmost portion
within the SPIRE footprint, insufficient for reliable aperture
photometry of the extended structure.
Near-infrared extinction mapping over the full 20~arcmin extent of the
extension yields $E(H-K) = 0.087 \pm 0.004$~mag, corresponding to
$A_V = 1.38$~mag and $N(\mathrm{H_2}) = 1.3 \times 10^{21}$~cm$^{-2}$.
As discussed in Section~\ref{sec:nirext}, the filamentary geometry
precludes a reliable total mass estimate from this aperture.
One Planck PGCC cold clump is detected within the extension:
G320.81$-$3.23, with a catalogued dust temperature of
$15.9 \pm 3.1$~K and a column density
$N(\mathrm{H_2}) = 3.80 \times 10^{20}$~cm$^{-2}$, yielding a clump
gas mass of approximately 65~M$_\odot$ at 750~pc.

\subsection{YSO Census}
\label{sec:yso}

\begin{figure}
\includegraphics[width=\columnwidth]{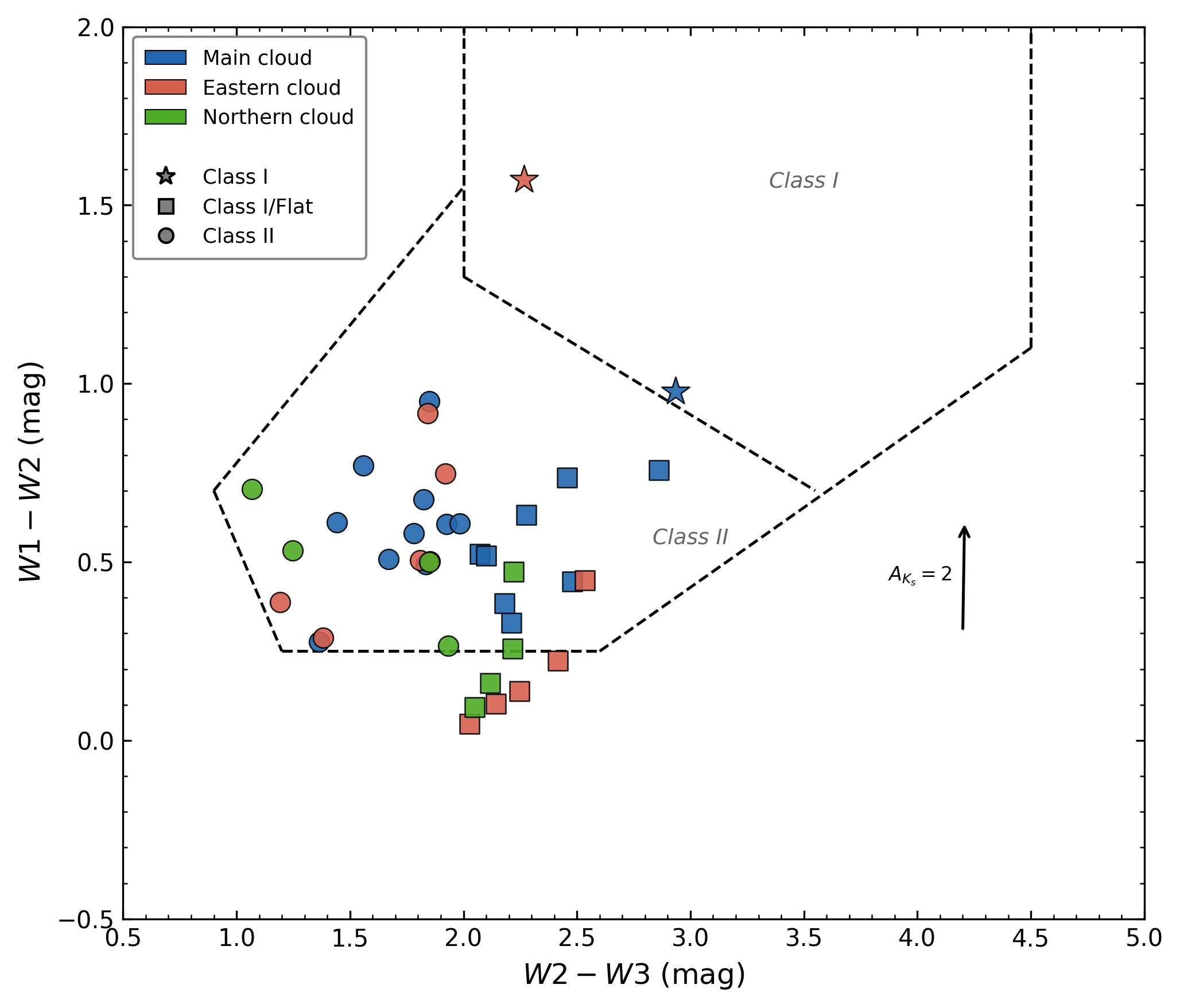}
\caption{AllWISE $W1-W2$ versus $W2-W3$ colour--colour diagrams for YSO 
candidates identified in the three components of the NE-Circinus Complex, 
classified using the scheme of \citet{Koenig2014}. Dashed lines show the K14 
classification boundaries. The $W3$ band photometry in the eastern 
component is affected by diffuse PAH emission; YSO counts for this component 
should be treated with slight caution.}
\label{fig:ccd}
\end{figure}

A total of 39~YSO candidates are identified across the three components
of the NE-Circinus Complex using the two-colour classification scheme of
\citet{Koenig2014}, summarised in Table~\ref{tab:yso} and shown in Figure~\ref{fig:ccd}.
The sample comprises 2~Class~I, 17~flat-spectrum, and 20~Class~II
sources.
Two sources satisfying the Class~II colour cuts were identified as
Mira variables via SIMBAD cross-matching --- IRAS~15259$-$6059
and OGLE~GD-LPV-8662 --- and were removed from the sample prior to
the counts reported here.

Residual contamination by extragalactic sources is estimated to be 
$\sim$13 per cent for Class~I candidates and $\lesssim$2 per cent for 
Class~II candidates, based on the simulations of \citet{Koenig2014} 
for equivalent Galactic latitudes. The reported YSO counts should be 
interpreted with this in mind, particularly for the Class~I sample 
which is small in absolute terms.
The complex is actively star-forming across all three morphological
components.

The Class~I source in the main cloud body is located at
RA\,$231.769\degr$, Dec\,$-60.985\degr$, coincident with a local
maximum in the SPIRE 250~$\mu$m emission.
The eastern component hosts a Class~I source at
RA\,$232.669\degr$, Dec\,$-60.848\degr$, notable given the low
optical detectability of this component in DSS2 Red imaging.
The northern filamentary extension hosts a YSO population broadly
consistent with its physical size relative to the other components,
though no Class~I sources are identified there.

At the adopted distance of 750~pc, the AllWISE YSO census is subject to 
greater distance-dependent incompleteness than equivalent mid-infrared 
surveys of nearby star-forming regions such as Lupus 
\citep[${\sim}150$--200~pc;][]{Gaczkowski2015}, and the 39 candidates 
should be treated as a lower bound on the true disc-bearing population.

\begin{figure*}
    \centering
    \includegraphics[width=\textwidth]{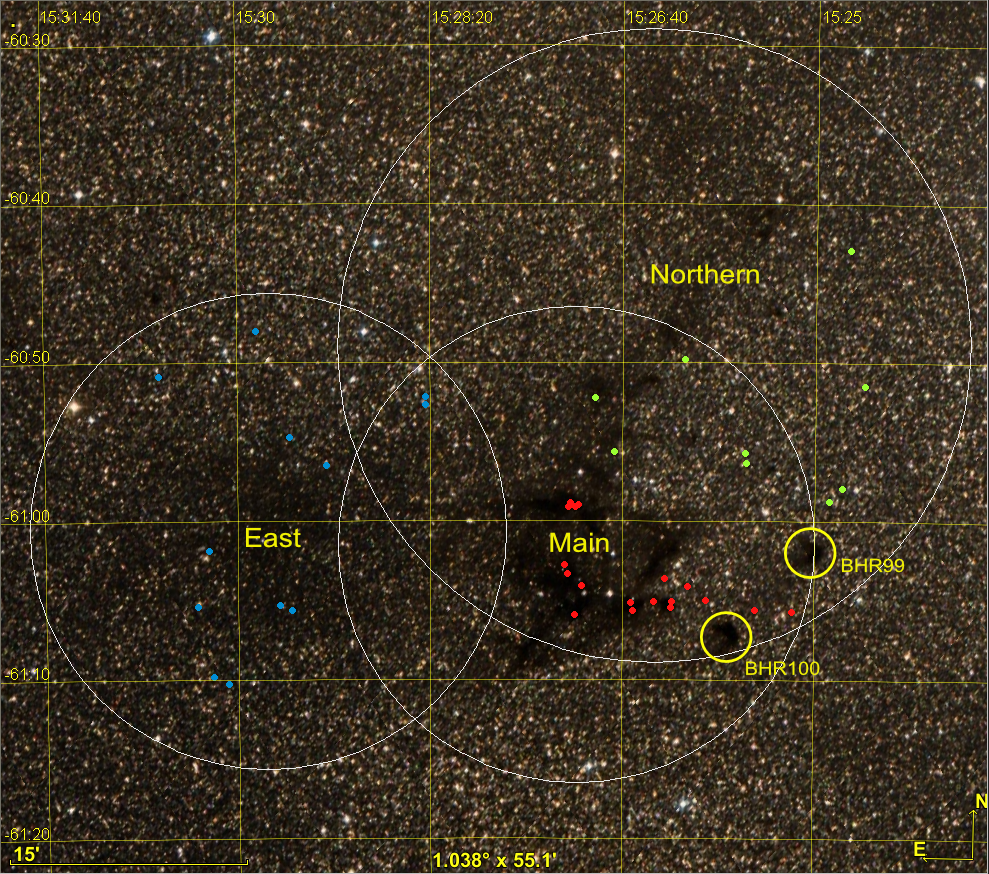}
    \caption{YSO candidates identified in the NE-Circinus Complex 
    overlaid on a DSS2 colour image. Red, blue, and green filled 
    circles indicate candidates associated with the main cloud body, 
    eastern component, and northern extension respectively, 
    classified using the scheme of \citet{Koenig2014}. White circles 
    show the AllWISE cone search areas for each component (15~arcmin 
    radius for the main body and eastern component; 20~arcmin radius 
    for the northern extension). Yellow circles mark the positional 
    exclusion regions around BHR~99 and BHR~100. North is up and 
    east is left.}
    \label{fig:yso_positions}
\end{figure*}

The positions and basic photometric properties of a representative 
sample of the identified YSO candidates are given in 
Table~\ref{tab:yso_positions}; the complete catalogue is available 
at \url{https://doi.org/10.5281/zenodo.19904310}.
The sky distribution of all 39~YSO candidates is shown in 
Figure~\ref{fig:yso_positions}, overlaid on the DSS2 image 
with candidates colour-coded by component.

\begin{table}
\centering
  \caption{YSO candidates identified in the NE-Circinus Complex from
    AllWISE photometry using the classification scheme of
    \citet{Koenig2014}.
    BHR~99 and BHR~100 are excluded by applying a positional exclusion
    radius of $0.05\degr$ around each globule.
    Two Mira variables identified via SIMBAD cross-matching have been
    removed from the eastern component count.}
  \label{tab:yso}
  \begin{tabular}{lrrrr}
    \hline
    Component & Class~I & Flat & Class~II & Total \\
    \hline
    Main cloud body        &  1 &  8 & 9 & 18 \\
    Eastern component$^a$  &  1 &  5 &  6 & 12 \\
    Northern extension     &  0 &  4 &  5 &  9 \\
    \hline
    Total                  &  2 & 17 & 20 & 39 \\
    \hline
\end{tabular}
  \begin{minipage}{\columnwidth}
    \smallskip
    $^a$\,$W3$ band photometry in this component is affected by diffuse
    PAH emission; counts should be treated with slight caution.
  \end{minipage}
\end{table}

\begin{table*}
  \centering
  \caption{Sample of YSO candidates identified in the NE-Circinus 
  Complex from AllWISE photometry, classified using the scheme of 
  \citet{Koenig2014}. Three sources from each component are shown; 
  the full 39-source catalogue is available at 
  \url{https://doi.org/10.5281/zenodo.19904310}.}
  \label{tab:yso_positions}
  \begin{tabular}{lrrrrrcl}
    \hline
    AllWISE & RA & Dec & $W1$ & $W2$ & $W3$ & Class & Component \\
            & (deg) & (deg) & (mag) & (mag) & (mag) & & \\
    \hline
    J152511.50-610542.8 & 231.2980 & -61.0952 & 10.130 & 9.637 & 7.803 & II & Main \\
    J152530.82-610535.7 & 231.3784 & -61.0933 & 12.536 & 12.090 & 9.611 & I/Flat & Main \\
    J152556.75-610457.8 & 231.4865 & -61.0827 & 11.147 & 10.625 & 8.553 & I/Flat & Main \\
    J152822.24-605239.0 & 232.0927 & -60.8775 & 6.328 & 5.411 & 3.568 & II & Eastern \\
    J152822.31-605211.7 & 232.0930 & -60.8699 & 12.067 & 11.929 & 9.682 & I/Flat & Eastern \\
    J152913.72-605630.3 & 232.3072 & -60.9418 & 14.264 & 14.040 & 11.624 & I/Flat & Eastern \\
    J152434.59-605126.8 & 231.1442 & -60.8575 & 12.703 & 12.438 & 10.506 & II & Northern \\
    J152442.45-604255.1 & 231.1769 & -60.7153 & 12.897 & 12.424 & 10.203 & I/Flat & Northern \\
    J152445.56-605754.2 & 231.1898 & -60.9651 & 13.790 & 13.533 & 11.317 & I/Flat & Northern \\
    \hline
  \end{tabular}
\end{table*}

\subsection{Infrared Variability}
\label{sec:variability}

\begin{figure}
    \includegraphics[width=\columnwidth]{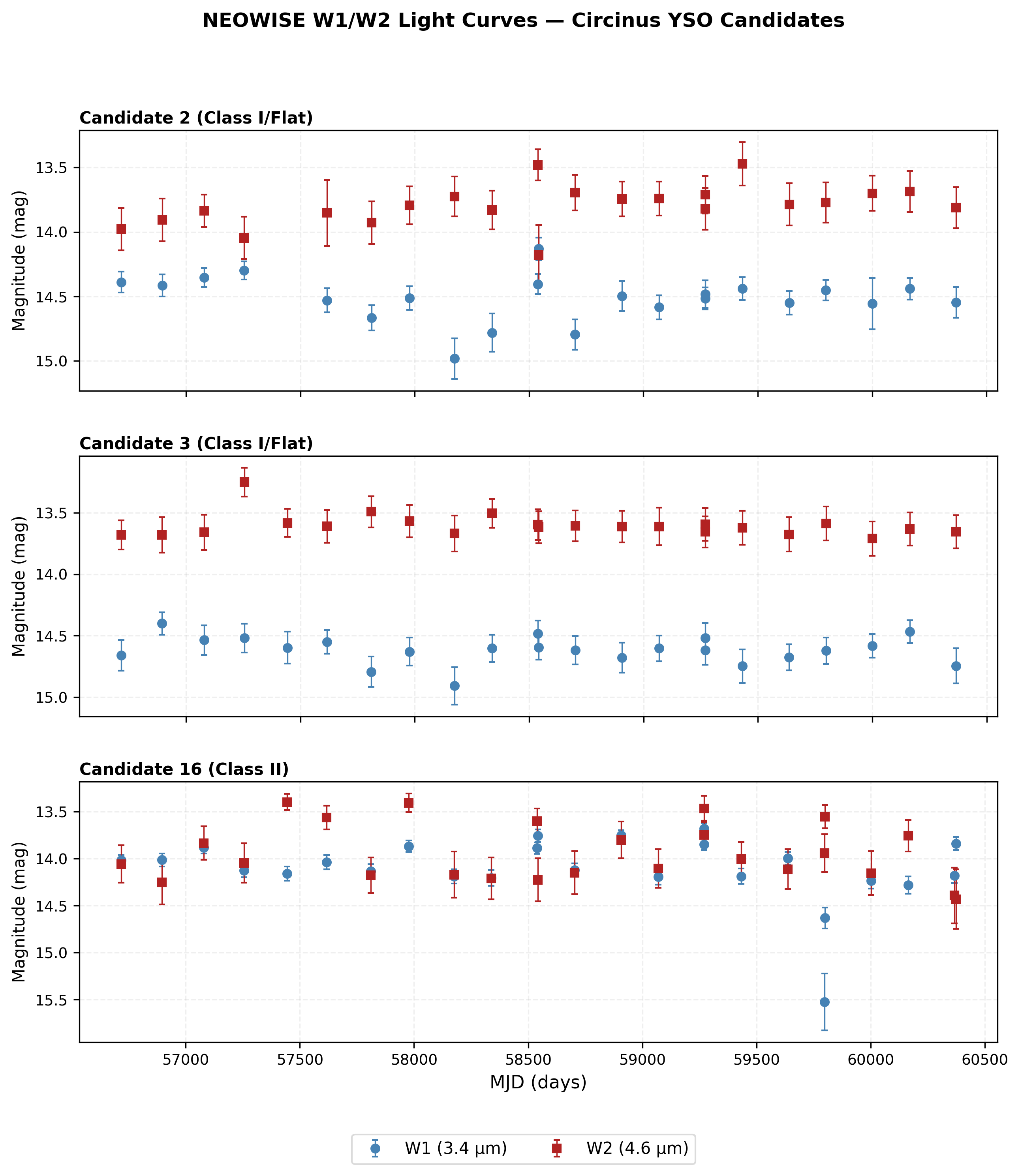}
    \caption{NEOWISE-R $W1$ (3.4~\micron, circles) and $W2$ (4.6~\micron,
    squares) light curves for the three significantly variable YSO candidates
    ($\Delta W1 > 0.5$~mag) in the NE-Circinus Complex. Error bars show the
    standard error on the per-epoch mean magnitude. Candidate~16 (Class~II,
    bottom panel) exhibits a pronounced aperiodic fading event near
    MJD~$\sim$59700, with the $W1$ fading substantially deeper than the
    contemporaneous $W2$ fading --- consistent with dust extinction rather than
    an accretion burst --- classifying this source as an aperiodic dipper.
    Candidates~2 and~3 (both Class~I/Flat, upper panels) show lower-amplitude
    correlated variability; both reach their deepest $W1$ fading at approximately
    MJD~57900.}
    \label{fig:lc}
\end{figure}

Of the 39~YSO candidates identified in Section~\ref{sec:yso}, 
NEOWISE-R light curves were able to be retrieved for 26~sources. Two of these 
--- the Mira variables IRAS~15259$-$6059 and OGLE~GD-LPV-8662 --- 
were removed from the variability sample following SIMBAD 
cross-matching (Section~\ref{sec:neowise}), leaving 24~candidates. 
A further three lacked sufficient epoch coverage for reliable 
variability analysis, yielding a clean sample of 21~candidates.
$W1$-band variability amplitudes ($\Delta W1 = \mathrm{max} - 
\mathrm{min}$) were computed for all 21 from the binned NEOWISE-R 
light curves.
A significance threshold of $\Delta W1 > 0.5$ mag was adopted to 
identify clearly variable sources. This threshold lies above the bulk 
of modest-amplitude YSO variability detected in NEOWISE surveys 
\citep{Park2021}, while remaining below the $\geq$1 mag amplitude cuts 
typically used to identify large-amplitude outbursting events, providing 
a balance between sensitivity and reliability for this sample size.
One further candidate within the clean sample exceeds this threshold but 
is excluded from the variable sample on the grounds of insufficient epoch 
coverage ($n_\mathrm{epoch} = 10$); its variability amplitude is therefore
unreliable.
Three sources with adequate epoch coverage exceed the threshold and
are listed in Table~\ref{tab:variable}; their $W1$ and $W2$ light
curves are shown in Figure~\ref{fig:lc}.

\begin{table}
\centering
  \caption{Significantly variable YSO candidates
    ($\Delta W1 > 0.5$~mag) identified from NEOWISE-R multi-epoch
    photometry of the NE-Circinus Complex.}
  \label{tab:variable}
  \begin{tabular}{lcrcc}
    \hline
    Candidate & Class & $\langle W1\rangle$ & $\Delta W1$ & $\Delta W2$ \\
              &       & (mag)               & (mag)       & (mag)       \\
    \hline
    2  & I/Flat &  14.52 & 0.85 & 0.71 \\
    3  & I/Flat &  14.62 & 0.51 & 0.46 \\
    16 & II     &  14.11 & 1.84 & 1.03 \\
    \hline
  \end{tabular}
\end{table}

The sole Class~I candidate has the faintest $W1$
magnitude in the sample ($W1 = 16.23$~mag, ph\_qual~B), consistent
with deeper embedding attenuating the shorter-wavelength band.

Candidate~16 (Class~II, $\Delta W1 = 1.84$~mag, $\Delta W2 = 1.03$~mag)
is the most strongly variable source in the sample.
Its light curve is dominated by a single pronounced fading event near
MJD~59700, in which the $W1$ flux drops to $\sim\!15.5$~mag before
recovering.
The $W1$ fading is substantially deeper than the contemporaneous $W2$
fading --- behaviour expected from dust extinction, which attenuates
the shorter-wavelength $W1$ band more strongly than $W2$.
The symmetric fading and recovery profile, the differential $W1/W2$
amplitude, and the absence of a colour change indicative of an
accretion burst classify this source as an aperiodic dipper (APD):
a YSO exhibiting dimming due to occultation of the stellar photosphere
by a warp or dust clump in the inner disc \citep{Cody2018}.
No SIMBAD counterpart was found for Candidate~16, supporting its
identification as a previously uncatalogued young stellar object.

Candidates~2 and~3 (both Class~I/Flat, $\Delta W1 = 0.85$ and
$0.51$~mag respectively) show lower-amplitude variability with
correlated $W1$ and $W2$ behaviour consistent with stochastic
extinction fluctuations typical of Class~I/Flat-spectrum protostars.
Candidate~3 is marginal at the adopted threshold.
Both sources show their deepest $W1$ fading at approximately
MJD~57900.

As NEOWISE visits the same sky position approximately every six months; 
the apparent simultaneity of the MJD $\sim$57900 fading in both sources 
could therefore reflect a single sky-pass rather than a physically 
correlated event.

Variability fractions were computed per class as $k/n$ with 95\%
Clopper--Pearson confidence intervals (Table~\ref{tab:varfrac}).
The Class~I/Flat fraction (2/11, 18.2\%) and Class~II fraction
(1/9, 11.1\%) are consistent with each other within the wide
confidence intervals reflecting the small sample size.
The Class~I sample ($n = 1$) is insufficient for a meaningful
variability fraction.

\begin{table}
    \centering
    \caption{Variability fractions per YSO class for the 21 clean
    NEOWISE-R candidates, with 95\% Clopper--Pearson confidence
    intervals.}
    \label{tab:varfrac}
    \begin{tabular}{lrrll}
    \hline
    Class  & $n$ & Variable & Fraction & 95\% CI \\
    \hline
    I      &  1 & 0 & 0.0\%  & [0.0\%, 97.5\%]  \\
    I/Flat & 11 & 2 & 18.2\% & [2.3\%, 51.8\%]  \\
    II     &  9 & 1 & 11.1\% & [0.3\%, 48.2\%]  \\
    \hline
    \end{tabular}
    
\end{table}
  
\section{Discussion}
\label{sec:discussion}

\subsection{A Heterogeneous Picture: Survey Coverage and Characterisation}
\label{sec:discussion_coverage}

The NE-Circinus Complex presents a striking example of the challenge of
characterising southern sky molecular clouds in the current era of
large-scale surveys.
No single dataset covers the entire complex at any wavelength.
The \textit{Herschel} SPIRE mosaic provides the most detailed
far-infrared characterisation available for the main cloud body, but
misses the eastern component entirely and captures only the southernmost
tip of the northern filamentary extension.
This is a direct consequence of the observational design --- the SPIRE
observations were targeted at the foreground Bok globules BHR~99 and
BHR~100, and the NE-Circinus Complex was identified serendipitously in
the resulting mosaic.
The eastern and northern components are therefore characterised by
near-infrared extinction mapping and Planck PGCC data alone,
supplemented by the AllWISE YSO census and the \citet{Guo2021} 3D
extinction map.

This heterogeneity reflects the general state of southern sky coverage
at far-infrared and submillimetre wavelengths rather than a weakness
unique to the present study.
A coherent physical picture can nonetheless be assembled from
incomplete, multi-facility archival data, combining \textit{Herschel},
\textit{Planck}, 2MASS, AllWISE, \textit{Gaia}, and ground-based
optical imaging to characterise a previously unstudied cloud complex
across all three of its morphological components.
This approach is directly applicable to other unstudied southern sky
clouds in similar situations.

\subsection{Cloud Mass and the Limits of Available Data}
\label{sec:discussion_mass}

The extinction-based gas masses derived in Section~\ref{sec:dust_main}
--- $\sim$277~M$_\odot$ within the dense 6~arcmin core and
$\sim$439~M$_\odot$ within the full 10~arcmin extent --- place the
NE-Circinus Complex firmly in the regime of modest low-mass
star-forming clouds.
Both values are robust to the kappa uncertainties that plague
submillimetre continuum mass estimates, since the near-infrared
colour excess method requires no assumed dust opacity.
The close agreement between the 6~arcmin extinction-derived column
density ($N(\mathrm{H_2}) = 2.3 \times 10^{21}$~cm$^{-2}$) and the
independently determined Planck PGCC value for G320.67$-$3.67
($2.36 \times 10^{21}$~cm$^{-2}$) provides strong cross-validation of
both measurements.

The NE-Circinus Complex, with a total gas mass of $\sim$439\,M$_\odot$ 
(10 arcmin aperture) and 39~YSO candidates, is broadly comparable in 
mass to the Lupus star-forming complex. The three main Lupus clouds 
(I, III, and IV) have a combined mass of $\sim$430\,M$_\odot$ above 
$A_V = 2$\,mag \citep{Benedettini2015} and host 38 YSOs/protostars 
detected with \textit{Herschel} \citep{Benedettini2018}. The Lupus 
complex lies at only $\sim$150--200\,pc \citep{Gaczkowski2015}, however, 
compared to 750\,pc for the NE-Circinus Complex. The AllWISE YSO census 
presented here is therefore subject to greater distance-dependent 
incompleteness than equivalent mid-infrared surveys of nearby clouds, 
and the 39 candidates should be treated as a lower bound on the true 
disc-bearing population.

A complete mass census of the entire complex would be unwieldy and 
highly uncertain based on current methods. The extinction-based mass 
of $\sim$439\,M$_\odot$ captures the dense main cloud body but 
excludes the northern extension and eastern component. Individual 
aperture masses for the discrete clumps of the northern extension 
are not feasible from 2MASS photometry, as the low background star 
density in this field ($\sim$2 stars per 4 arcmin$^2$) is 
insufficient for reliable colour excess measurements at sub-arcminute 
scales. The PGCC clump mass of $\sim$65\,M$_\odot$ for 
G320.81$-$3.23 therefore provides the best available lower bound on 
the dense gas content of the northern extension, giving a combined 
minimum mass for the complex of $\sim$500\,M$_\odot$.

The SPIRE data provide a dust temperature of $14.5 \pm 0.5$~K for the
main cloud body but cannot be used to derive a reliable total mass
without a Planck-based zero-point correction to the absolute surface
brightness \citep{Sadavoy2018}.
The SPIRE-derived temperature of $14.5 \pm 0.5$\,K and the 
Planck PGCC temperature of $13.7 \pm 0.9$\,K for 
G320.67$-$3.67 are consistent within their uncertainties. 
The small systematic offset, in which the SPIRE value is 
marginally warmer, is expected from beam dilution: the 
Planck beam at $\sim$4.5\,arcmin FWHM averages over a 
larger solid angle than the SPIRE measurement, incorporating 
cooler outer envelope material and producing a systematically 
lower beam-averaged temperature.
This beam dilution effect is a recurring interpretive theme for the
eastern and northern components, where Planck-derived properties
represent beam-averaged quantities that may differ significantly from
the true peak conditions.

The near-infrared extinction method is well-suited to the main cloud
body, where the star density is sufficient and the cloud geometry is
approximately circular.
For the eastern component, the weak extinction signal
($A_V = 0.49$~mag) is consistent with its diffuse optical morphology
but too uncertain for a reliable mass estimate.
For the northern filamentary extension, the elongated geometry is
poorly matched by a circular aperture and the mean extinction is
heavily diluted by the large fraction of low-extinction sky enclosed
within the 20~arcmin search area; resolved extinction mapping along
the filament spine would be required for a meaningful mass estimate.
The PGCC clump mass of $\sim$65~M$_\odot$ for G320.81$-$3.23 provides
a lower bound on the dense gas content of the extension.

\subsection{Relationship to the Circinus Complex}
\label{sec:discussion_circinus}

The NE-Circinus Complex lies in proximity to the broader Circinus 
star-forming region, centred on the well-studied Cir-E and Cir-W 
molecular clouds, which has been the subject of comprehensive 
stellar population analysis by \citet{Kerr2025}. The NE-Circinus 
Complex lies approximately $2\degr$ to the north-east of Cir-E 
in projection, and is clearly distinct from those clouds both 
morphologically and spatially. \citet{Kerr2025} note the existence 
of an additional molecular gas cloud hosting potential dense cores 
located to the Galactic north of Cir-E and Cir-W, previously 
identified in extinction catalogues but assigned no name and 
receiving no dedicated study \citep{dobashi2005}. We identify 
the NE-Circinus Complex with this unnamed structure.

That survey also identified the stellar subgroup OC-0625, with a 
mean position of RA $231.7\degr$, Dec $-61.1\degr$, lying within 
the projected boundary of the NE-Circinus Complex main cloud body, 
and a mean distance of 821\,pc broadly consistent with our adopted 
distance of $750 \pm 50$\,pc. This suggests a possible physical 
association, in which case OC-0625 may represent a population of 
young stars that have partially dispersed from the gas complex. 
Kinematic data --- molecular line radial velocities and stellar 
proper motions --- would be needed to confirm this association.

The most recent dedicated study of the broader Circinus 
star-forming region, \citet{Rector2025}, surveyed Herbig-Haro 
objects across Cir-W and Cir-E using wide-field DECam imaging, 
but the NE-Circinus Complex falls outside their survey footprint 
entirely, further confirming that this structure has received no 
dedicated attention in the published literature prior to the 
present work.

\subsection{The Eastern Component}
\label{sec:discussion_eastern}

Despite its low optical detectability and the absence of
\textit{Herschel} SPIRE coverage, the eastern component hosts
12~YSO candidates --- a surface density comparable to, and in relative
terms consistent with, that of the optically more prominent main cloud body.
The relatively high ratio of Class~II to Class~I sources in the eastern
component compared to the main cloud body may tentatively suggest a
somewhat more evolved star-forming population, in which a larger
fraction of young stars have dispersed their inner disc material,
consistent with the eastern component being either older or more
diffuse in nature than the main cloud body.
Caution should be taken however, as $W3$ photometry in this component is affected
by diffuse PAH emission.

While the \citet{Koenig2014} classification cuts are designed to
mitigate contamination from spurious $W3$ detections, PAH emission is
a known source of residual contamination in this band, and the eastern
component YSO counts should be treated with appropriate caution pending
higher-resolution mid-infrared observations.

Cross-matching the eastern component YSO candidates against Gaia DR3 
within 2 arcsec yields no counterparts, precluding a proper motion 
membership test with current data. The absence of Gaia detections is 
consistent with the intrinsically red spectral energy distributions 
of disc-bearing young stars, which are typically faint at optical 
wavelengths. Kinematic confirmation of eastern component membership 
awaits future dedicated near-infrared astrometric observations.

\subsection{The Northern Filamentary Extension and Uncatalogued Globules}
\label{sec:discussion_northern}

The northern filamentary extension presents a morphology consistent
with the filamentary star-forming structures described by
\citet{Andre2014} --- dense and well-defined in its southern portion,
thinning and dissolving progressively northward over a projected length
of approximately 3.5~pc.
Its YSO population of 9~candidates confirms ongoing star formation
within the extension.

Of particular note are at least two compact, apparently uncatalogued Bok
globules visible on the western edge of the northern extension, the two most prominent listed
in Table~\ref{tab:globules}.
These objects are distinct from the foreground Bok globules BHR~99 and
BHR~100, which lie at $\sim$350~pc and are physically unassociated
with the NE-Circinus Complex. Whether these globules are physically associated with the 
NE-Circinus Complex or are foreground or background objects 
projected against it remains to be determined; molecular line observations would establish a common systemic radial velocity and resolve this question.
They  remain absent from SIMBAD and all major southern dark cloud catalogues, including the BHR catalogue \citep{Bourke1995} and the DC catalogue.

Their compact morphology and location at the edge of an active
star-forming filament make them strong candidates for dense starless or
protostellar cores.

Both NE-Cir Globule 1 and Globule 2 have compact counterparts 
detected in the Herschel SPIRE 250\,$\mu$m mosaic at signal-to-noise 
ratios of $\sim$22 and $\sim$35 respectively, and are detected at 
350 and 500\,$\mu$m with comparable significance. Globule 1 lies 
near the northern mosaic edge where coverage depth is reduced, but 
remains clearly detected at all three SPIRE wavelengths. Modified 
blackbody fits with $\beta = 2$ fixed to the 250, 350, and 
500\,$\mu$m flux densities, measured from maps convolved to a common 
resolution of 36.6 arcsec (the 500\,$\mu$m beam), yield dust 
temperatures of $15.2 \pm 0.8$\,K and $13.5 \pm 0.4$\,K for 
Globule 1 and Globule 2 respectively. These temperatures are 
consistent with the cold dense material characterising the northern 
extension (PGCC G320.81$-$3.23: $15.9 \pm 3.1$\,K) and the main 
cloud body ($14.5 \pm 0.5$\,K), supporting a physical association 
with the NE-Circinus Complex at $\sim$750\,pc. The quoted 
uncertainties are formal fitting errors only; systematic 
uncertainties from background subtraction and the assumed $\beta$ 
contribute an additional $\sim$2--3\,K. Both globules are strong 
candidates for dense starless or protostellar cores and represent 
the highest-priority targets for dedicated follow-up within the 
complex.

Molecular line observations, particularly CO and its isotopologues, would 
establish whether the globules are kinematically associated with the 
northern extension and constrain their internal velocity structure; 
high-resolution ALMA continuum imaging at 1.3~mm \citep{Wootten2009} would determine whether 
either harbours a compact protostellar source.

\begin{table}
\centering
\caption{Uncatalogued compact Bok globules on the western edge of the
northern filamentary extension. Dust temperatures from modified blackbody
fits ($\beta = 2$ fixed) to SPIRE 250, 350, and 500\,$\mu$m photometry
convolved to a common 36.6 arcsec resolution.}
\label{tab:globules}
\begin{tabular}{lccc}
\hline
Globule & $l$ ($\degr$) & $b$ ($\degr$) & $T_\mathrm{dust}$ (K) \\
\hline
NE-Cir Globule 1 & 320.648 & $-$3.315 & $15.2 \pm 0.8^{a}$ \\
NE-Cir Globule 2 & 320.560 & $-$3.455 & $13.5 \pm 0.4$ \\
\hline
\multicolumn{4}{l}{$^{a}$ Near mosaic edge; background estimate one-sided.}\\
\end{tabular}
\end{table}

\subsection{Comparison with Sadavoy et al.}
\label{sec:discussion_sadavoy}

\citet{Sadavoy2018} used the same \textit{Herschel} SPIRE observations
in a multi-cloud study of Bok globule dust properties, producing dust
temperature and optical depth maps for BHR~99 and BHR~100 as part of a
sample of 56 southern globules.
The NE-Circinus Complex, though clearly present as a bright, extended
far-infrared source in the SPIRE mosaic, was not identified as a
distinct object of study in that work --- unsurprisingly, since the
observations were designed to target the compact foreground globules,
and the background cloud complex appears as extended emission at the
edge of the field rather than as a primary science target.

The present paper demonstrates that archival \textit{Herschel} data can
harbour significant unreported science beyond its original observational
intent.
The NE-Circinus Complex was identified not through a dedicated
observational programme but through serendipitous coverage of a
background cloud in observations targeting unrelated foreground objects.
That a previously unstudied, actively star-forming cloud complex had
remained uncharacterised underscores both the richness of the \textit{Herschel} 
archive and the incomplete state of southern sky molecular cloud inventories, and raises
the question of how many similar structures remain unidentified in
archival data targeting other southern sky Bok globule fields.

\subsection{Infrared Variability and the Disc-Bearing Population}
\label{sec:discussion_variability}

The overall variability fractions derived from NEOWISE-R photometry are
low, consistent with the youth and embeddedness of the NE-Circinus
Complex population.
NEOWISE-R variability detection is inherently incomplete for deeply
embedded sources: the sole Class~I candidate in the sample has the
faintest $W1$ magnitude (16.23~mag) and poorest $W1$ quality flag of
any source, and its non-detection as a variable source likely reflects
sensitivity limitations at this evolutionary stage rather than a
physical absence of variability.
The variability fractions should therefore be treated as lower limits
on the true variable fraction.

The disc-bearing population is more completely characterised by the
AllWISE colour selection itself.
All 21~candidates with usable NEOWISE-R light curves retain their
Class~I, I/Flat, or II classifications, confirming the presence of
infrared excess and therefore circumstellar disc or envelope material
regardless of detected variability.
The absence of a complete stellar census for the NE-Circinus Complex
--- in particular the Class~III population, which is undetectable by
WISE colour selection alone --- precludes calculation of an overall
disc fraction for the complex.

Candidate~16 is the most notable individual result of the variability
analysis.
Its large $W1$ amplitude ($\Delta W1 = 1.84$~mag), the differential
depth of the fading event between $W1$ and $W2$ consistent with dust
extinction, the symmetric light curve morphology, and the absence of
any prior catalogue identification collectively support its
classification as an aperiodic dipper.
Dippers are understood to arise from occultation of the stellar
photosphere by warps or dust clumps in the inner disc, typically seen
at disc inclinations $i > 50\degr$, and are physically distinct from
eruptive variables in which accretion outbursts drive brightening
rather than fading \citep{Cody2018}.
Candidate~16 represents the most specific characterisation of an
individual young stellar object within the NE-Circinus Complex to date
and warrants follow-up near-infrared monitoring to characterise whether
the fading events recur and to constrain the disc geometry.  
  
\section{Conclusions}
\label{sec:conclusions}

This paper presents the first dedicated characterisation of the
NE-Circinus Complex (TGU~H1984\,/\,DCld\,320.7$-$03.6), a previously
unstudied star-forming dark cloud complex serendipitously identified in
archival \textit{Herschel} SPIRE data at a distance of $750 \pm 50$~pc.
The complex consists of three morphologically distinct components whose
physical properties have been characterised through a combination of
\textit{Herschel} SPIRE photometry, near-infrared 2MASS colour excess
mapping, Planck PGCC catalogued clump properties, AllWISE YSO
classification, and 3D dust extinction mapping.
Our main conclusions are as follows.

\begin{enumerate}

\item The NE-Circinus Complex is actively star-forming across all three
  of its components, hosting a total of 39~YSO candidates identified
  from AllWISE photometry using the \citet{Koenig2014} classification
  scheme.
  That an actively star-forming complex of this scale had gone entirely
  uncharacterised in the published literature --- despite lying within
  the well-studied Circinus star-forming region and being present in
  publicly available \textit{Herschel} data since 2013 --- reflects
  the general incompleteness of southern sky molecular cloud inventories
  and the heterogeneous nature of far-infrared survey coverage at these
  Galactic latitudes.

\item The main cloud body has a SPIRE-derived dust temperature of
  $14.5 \pm 0.5$~K.
  Near-infrared $H-K$ colour excess mapping of background 2MASS stars
  yields a core gas mass of $\sim$277~M$_\odot$ within a 6~arcmin
  aperture (physical radius 1.3~pc) and a total gas mass of
  $\sim$439~M$_\odot$ within a 10~arcmin aperture (2.2~pc), both at
  750~pc.
  The 6~arcmin column density ($N(\mathrm{H_2}) = 2.3 \times
  10^{21}$~cm$^{-2}$) is in close agreement with the independently
  determined Planck PGCC value for G320.67$-$3.67
  ($2.36 \times 10^{21}$~cm$^{-2}$), providing strong cross-validation
  of the extinction measurement.
  Reliable total envelope mass characterisation from SPIRE continuum
  data would require a Planck-based zero-point correction to the
  absolute surface brightness \citep{Sadavoy2018}.

\item Multi-epoch NEOWISE-R photometry of the YSO candidate population
  reveals three significantly variable sources.
  The most notable, Candidate~16 (Class~II), exhibits a large-amplitude
  aperiodic fading event with differential $W1/W2$ extinction consistent
  with inner-disc occultation, classifying it as an aperiodic dipper
  \citep{Cody2018}.
  Candidate~16 is absent from all prior catalogues and represents the
  most detailed characterisation of an individual young stellar object
  within the complex to date.

\item Two uncatalogued compact Bok globules identified on the western edge 
of the northern filamentary extension are absent from all major 
southern dark cloud catalogues. Both are detected in Herschel SPIRE 
continuum emission, with modified blackbody dust temperatures of 
$15.2 \pm 0.8$\,K (NE-Cir Globule~1) and $13.5 \pm 0.4$\,K 
(NE-Cir Globule~2).

\item A possible physical association between the NE-Circinus Complex
  and the OC-0625 stellar subgroup of \citet{Kerr2025} is suggested by
  their positional coincidence and broadly consistent distances, but
  cannot be confirmed without molecular line observations providing a
  systemic gas radial velocity for the complex.

\item The serendipitous discovery of the NE-Circinus Complex in
  observations targeting unrelated foreground objects demonstrates the
  scientific value of systematic examination of archival
  \textit{Herschel} data beyond its original observational intent, and
  motivates similar searches in other southern sky Bok globule fields.

\end{enumerate}

\section*{Acknowledgements}

The \textit{Herschel} Space Observatory observations used in this work 
were obtained under Open Time programme OT2\_tbourke\_3 (PI: T.~Bourke). 
\textit{Herschel} is an ESA space observatory with science instruments 
provided by European-led Principal Investigator consortia and with 
important participation from NASA.

This work has made use of data from the European Space Agency (ESA) 
mission \textit{Gaia} (\url{https://www.cosmos.esa.int/gaia}), processed 
by the \textit{Gaia} Data Processing and Analysis Consortium (DPAC, 
\url{https://www.cosmos.esa.int/web/gaia/dpac/consortium}). Funding for 
the DPAC has been provided by national institutions, in particular the 
institutions participating in the \textit{Gaia} Multilateral Agreement.

This publication makes use of data products from the Wide-field Infrared 
Survey Explorer, which is a joint project of the University of California, 
Los Angeles, and the Jet Propulsion Laboratory/California Institute of 
Technology, funded by the National Aeronautics and Space Administration.

This research has made use of the VizieR catalogue access tool, CDS, 
Strasbourg, France \citep{Ochsenbein2000}, and the SIMBAD database, 
operated at CDS, Strasbourg, France.

This research has made use of \textsc{topcat} \citep{Taylor2005}.

This research made use of Claude AI (Anthropic) as a programming assistant for the generation and debugging of Python scripts used in figure production and data analysis. It was also used for determination of relevant literature, which was followed up by the author for validation and analysis. All scientific interpretation, data analysis decisions, and conclusions are solely those of the author. All 
AI-assisted content was reviewed and verified by the author prior to inclusion.

\section*{Data Availability}

The \textit{Herschel} SPIRE observations used in this work are available 
from the Herschel Science Archive under observation IDs 1342265318 and 
1342265319. AllWISE photometry is available via the VizieR catalogue 
service. NEOWISE-R single-exposure photometry is available from the NASA 
IRSA archive. Gaia~DR3 data are available from the ESA Gaia archive. The 
Planck PGCC catalogue is available via VizieR as J/A\&A/594/A28/pgcc. 
Processed data products derived in this work, including the AllWISE YSO 
candidate catalogue with classifications and the NEOWISE-R binned light 
curves, are available at \url{https://doi.org/10.5281/zenodo.19904310}.



\bibliographystyle{mnras}
\bibliography{bibliography} 




\bsp	
\label{lastpage}
\end{document}